\documentclass[12pt]{article} 
\usepackage{titlesec} 
\usepackage{placeins}
\usepackage{units} 
\usepackage[sharp]{easylist} 
\usepackage{array,booktabs}
\usepackage{colortbl}
\usepackage{abbrevs}
\usepackage{etoolbox}
\usepackage{booktabs}  
\usepackage{amsmath,amssymb,amsfonts}  
\usepackage{textcomp} 
\usepackage{lipsum}   
\usepackage{dcolumn}
\usepackage{bm}
\usepackage{cancel}
\usepackage{verbatim}
\usepackage{ifthen}
\usepackage{url}  
\usepackage{sectsty}
\usepackage{balance}  
\usepackage{graphicx}
\usepackage{lastpage}
\usepackage[format=plain,justification=RaggedRight,singlelinecheck=false,
font=small,labelfont=bf,labelsep=period]{caption} 
\usepackage{fancyhdr}
\usepackage{color}
\usepackage{calc}
\usepackage{pdflscape}
\usepackage[utf8]{inputenc}
\usepackage[english]{babel}
\usepackage{natbib}
\usepackage{makecell}
\usepackage{comment}   
\usepackage{verbatim} 
\usepackage{lipsum, setspace} 
\usepackage{multirow}
\usepackage{pdfpages}
\usepackage{pdflscape}
\usepackage{pifont}
\usepackage{units}
\usepackage{mathtools}
\usepackage{dsfont}  
\usepackage{tcolorbox}
\usepackage[font=small,labelfont=bf]{caption} 
\usepackage{enumerate}
\usepackage{blindtext} 
\usepackage{tikz}
\usetikzlibrary{calc}
\usepackage[english]{babel} 
\usepackage{subcaption}   
\usepackage{caption}
\usepackage{setspace}
\usepackage{fullpage}
\usepackage[hidelinks]{hyperref}

\newlength{\chaptercapitalheight}
\newlength{\chapterfootskip}
\newlength\graphht

\fancypagestyle{lscapedplain}{%
  \fancyhf{}
  \fancyfoot{
    \tikz[remember picture,overlay]
      \node[outer sep=1cm,above,rotate=90] at (current page.east) {\thepage};}

}

\newcommand{\beq}{ \begin{equation}}
\newcommand{\eeq}{ \end{equation}}  
\newcommand{\beqn}{ \begin{eqnarray}}
\newcommand{\eeqn}{ \end{eqnarray}}

\newcommand{\bX}{\mathbf{X}}

\newcommand{\eps}{\epsilon}

\newcommand{\1}{\mathbf{1}}

\newcommand{\trips}[1]{{\left\vert\kern-0.25ex\left\vert\kern-0.25ex\left\vert #1
        \right\vert\kern-0.25ex\right\vert\kern-0.25ex\right\vert}}

\DeclareMathOperator{\E}{E}

\newtheorem{assumption}{Assumption}              

\newcolumntype{L}{@{}>{\kern\tabcolsep}l<{\kern\tabcolsep}}
\begin{document}

\vspace*{.6in}
\begin{center}
{\Large \textbf{Instrumental variables, spatial confounding and}}

\vspace{8pt}
{\Large \textbf{interference}}
\\ \vspace{6pt}

\vspace*{.3in}
{\large Andrew Giffin\footnote[1]{North Carolina State University, Department of Statistics}, Brian J.~Reich$^1$, Shu Yang$^1$, Ana G. Rappold\footnote[2]{Environmental Protection Agency}}\\
\vspace*{.1in}

\today
\end{center}

\vspace{.1in}

\begin{abstract}\begin{singlespace}
\noindent Unobserved spatial confounding variables are prevalent in environmental and ecological applications where the system under study is complex and the data are often observational. Instrumental variables (IVs) are a common way to address unobserved confounding; however, the efficacy of using IVs on spatial confounding is largely unknown. This paper explores the effectiveness of IVs in this situation -- with particular attention paid to the spatial scale of the instrument. We show that, in case of spatially-dependent treatments,  IVs are most effective when they vary at a finer spatial resolution than the treatment. We investigate IV performance in extensive simulations and apply the model in the example of long term trends in the air pollution and  cardiovascular mortality in the United States over 1990-2010. Finally, the IV approach is also extended to the spatial interference setting, in which treatments can affect nearby responses. \vspace{12pt} \end{singlespace}\end{abstract}

\vspace{.1in}

\section{Introduction}

Unobserved spatial confounding is a very common issue among observational studies in environmental and ecological applications. For example, the causal effect of air pollution on cardiovascular mortality is likely confounded by a number of spatially-dependent health and socioeconomic variables which may not be available to the researcher. Spatial confounding exists when both a treatment/exposure variable as well as a response variable are both related to an unobserved spatially varying confounder. In this situation, confounding can often be mitigated by imposing a spatial form as a latent process or incorporating spatial error. However, incorporating spatial error to a model does not necessarily address the bias due to unmeasured confounding \citep{hodges2010adding,reich2020review}. A number of other techniques have been shown to be effective as well. \cite{papadogeorgou2019adjusting,keller2020selecting, davis2019addressing, schnell2019mitigating, thaden2018structural, jarner2002estimation} all propose methods to reduce bias from spatial confounders.  

\cite{paciorek2010importance} shows that spatial confounding can be removed using splines, but only when the exposure variability has smaller spatial range than that of the confounder. This is a key finding that will be mirrored in our analysis. Our goal, however, is to mitigate this unobserved spatial confounding using an instrumental variable (IV). 

Instrumental variable methods provide a means of obtaining causal estimates even when confounders are unobserved, so long as a valid instrument exists. There is a sizable literature, primarily in the fields of economics and econometrics, that deals with the intersection of the IV approach and spatial models. Generally the models estimated are Spatial Autoregressive (SAR) models, which allow for dependence in the response as determined by a spatial weights matrix \citep{kelejian2004instrumental}. \cite{betz2017spatial} investigate these SAR models under spatial confounding, and show that if the response is spatially dependent, naive estimates can be biased. Moreover, if the IV is spatially correlated, this bias is worsened \citep{betz2017spatial}. The SAR/IV literature is relatively mature: the SAR/IV approach has been extended to quantile regression \citep{su2007instrumental}, endogenous spatial distance matrices \citep{qu2015estimating}, and longitudinal data models \citep{kelejian2014estimation, qu2016instrumental}. One popular extension to SAR models is the so-called SARAR model, which allows for spatial errors -- that is, SAR errors. This model has also received IV treatment \citep{kelejian1998generalized,kelejian2004instrumental, piras2010sphet, lee2003best}.  These various forms of SAR models are specified differently from the standard geostatistical or Conditional Autoregressive (CAR) models preferred in the statistics literature; however, they are broadly similar \citep{ver2018relationship}. 

This paper examines the efficacy of using both IVs that exhibit spatial dependence (henceforth, ``spatial'' IVs) as well as IVs that exhibit only weak spatial dependence (henceforth, ``local'' or ``non-spatial'' IVs) in the presence of spatially varying treatment and unmeasured confounding. We find that local instrumental variables tend to provide more information with which to remove confounding bias, than those with spatial dependence. This suggests that local instruments are more effective at reducing confounding and isolating causal effects of treatment. This is seen in both reduced bias and variance when compared to spatial instruments. After Section \ref{s3:prelims} establishes the method and assumptions used, Section \ref{s3:sim} details the behavior of spatial and non-spatial IVs under different scenarios, using a series of simulation studies to compare and contrast spatial versus non-spatial instruments. Section \ref{s3:realData} provides a demonstration which estimates the causal effect of PM$_{2.5}$ on cardiovascular mortality in the United States over 1990-2010, using local emission levels as an instrument. Section \ref{s3:introInterference} extends the method to the spatial interference setting, and provides a simulation study to validate the proposed approaches (Section \ref{s3:simInterference}). Section \ref{s3:discussion} concludes with a discussion of the advantages and limitations of the method.

\section{Local versus spatial IVs and spatial confounding}\label{s3:prelims}

\subsection{Potential outcomes and instrumental variables}

We consider spatial processes in area of interest $\mathcal{D} \subset \mathbb{R}^2$, in which we observe data at $n$ locations $\mathcal{S} = \{s_1, \ldots, s_n\} \subset \mathcal{D}$.  At each of these points $s$ we observe a treatment $A_s \in \mathbb{R}^1$, a response $Y_s \in \mathbb{R}^1$, a covariate vector $\bX_s \in \mathbb{R}^p$, and an instrumental variable $Z_s \in \mathbb{R}^1$. Variables without a subscript denote entire fields, e.g., $A \equiv \{ A_s: s \in \mathcal{D}\}$. Under-case variables denote realized values; and subscript $_{-s}$ denotes all points in $\cal S$ except $s$.

We use the potential outcomes framework to define the causal effect \citep{rubin1974estimating}. We first consider a no interference situation: For each treatment at location $s$, the potential treatment that would be observed under instrument $z_s$ is written as $A_s(z_s)$; the potential outcome that would be observed under treatment $a_s$ is written as $Y_s(a_s)$. The temporal order has the instrument occurring before the treatment,  followed by the response. In addition, we invoke several key assumptions that will allow us to isolate causal effects. 

\begin{assumption}[No interference] \label{a:noInterference}
The potential outcome $Y_s(a)$ depends only on $a_s$, and not the treatments at other locations $a_{-s}$.  That is, $Y_s(a) = Y_s(a_s)$.
\end{assumption}

\begin{assumption}[Marginal Structural Model]
The potential outcomes for $A(z)$ and $Y(a)$ take the following form:
\begin{align}
  Y_s(a_s) &= \beta_0 + \delta_1 a_s + U_s + \epsilon_{1,s}, \label{eq:YaformNoInt}\\
    A_s(z_s) &= \gamma_0 + \delta_3 z_s + \gamma_1 U_s + \epsilon_{2,s},\label{eq:AzformNoInt}
\end{align}
where $U$ is an unobserved, potentially spatially-dependent confounder with mean zero, and $\epsilon_1$ and $\epsilon_2$ are mean zero error processes. Both error terms are potentially spatially dependent. We omit covariates in the above for clarity; however, we note that covariates $\bX_s$ can be accounted for by including them on the right-hand-side in both (\ref{eq:YaformNoInt}) and (\ref{eq:AzformNoInt}). (We reserve $\delta_2$ as the coefficient for the indirect effect in the presence of interference in Section \ref{s3:introInterference}.)
\end{assumption}

\begin{assumption}[Valid instrumental variable]\label{a:validIV}
$Z$ is independent of $\epsilon_1$, $\epsilon_2$, and $U$.  
\end{assumption}

\begin{assumption}[Consistency] 
The potential outcome $Y_s(a)$ is the equal to the response $Y_s$ that would be observed under treatment $A_s$. Similarly, $A_s(z)$ is the equal to the treatment $A_s$ that would be observed under instrument $z$.
\end{assumption}

The marginal structural model gives the form of the potential outcomes, and illustrates that the parameter $\delta_1$ will encapsulate the average treatment effect of $A$ on $Y$. Assumption \ref{a:validIV} ensures that $Z$ is uncorrelated with $U$. The consistency and no-interference assumptions ensure that our potential outcomes line up with the observed outcomes, and that it is enough to condition on the local treatment. We will relax Assumption \ref{a:noInterference} later in Section \ref{s3:introInterference} to investigate the method when interference is present.

\subsection{Identification}

This section establishes that with a valid instrument $Z$, we can recover the true treatment effect $\delta_1$ even when $A$ and $Y$ are confounded by an unmeasured confounder $U$. 
We can take the conditional expectation of (\ref{eq:YaformNoInt}), substitute $A_s$ for $a_s$, and write
\begin{align*}
\E [ Y_s(A_s) ~|~ Z_s] 
&= \E [ \beta_0 + \delta_1 A_s + U_s + \epsilon_{1,s} ~|~ Z_s ] \\
&= \beta_0 + \delta_1 \E [A_s|Z_s] + \E [U_s|Z_s] + \E[\epsilon_{1,s} | Z_s] \\
&=  \beta_0 + \delta_1 \E [A_s|Z_s]  \\
&=  \beta_0 + \delta_1 \hat{A}_s(Z_s).
\end{align*}
where $\hat{A}_s(Z) = \E [ A_s | Z]$. We will henceforth denote $\hat{A}_s(Z)$ by $\hat{A}$ for simplicity.

The first equality follows from (\ref{eq:YaformNoInt}), and the second from Assumption \ref{a:validIV}. From this we see that a regression from the observed $Y_s$ onto $\hat{A}_s$ still provides an unbiased estimator of the causal parameter value $\delta_1$. Assuming a linear relationship between $A_s$ and $Z_s$, we can approximate $\hat{A}_s$ with the fitted values from $A_s$ onto $Z$, which will fulfill $\hat{A}_s = \E [A_s|Z]$. A comprehensive primer on the fundamentals of IVs and potential outcomes is given in  \cite{angrist1996identification}.

\subsection{Estimation}\label{ss3:estimation}

We use a 2-stage estimation procedure which generalizes the 2-stage least squares (2SLS) of  \cite{wright1928tariff} and \cite{theil1958economic}. In the first stage, $A_s$ is regressed onto $Z_s$ via the regression $$A_s = \gamma_0 + \delta_3 Z_s + e_{s,1}$$ where the errors $e_{s,1}$ can either be spatially-correlated or independent. We then set $\hat{A}_s = \hat{\gamma}_0 + \hat{\delta}_3 Z_s$. In the second stage, $Y_s$ is regressed onto $\hat{A}_s$ via $$Y_s = \beta_0 + \delta_1 \hat{A}_s + e_{2,s}$$ with either spatially-correlated or independent $e_{2,s}$.  The estimate of $\delta_1$ from the second regression is our estimate of the causal effect. 

The spatial-error regressions are estimated with the R package geoR \citep{geoR}. In both stages, the spatial error model is a Gaussian process with mean zero and isotropic exponential covariance. The causal estimator is taken to be the coefficient point estimate for $\hat{A}$ from the second stage regression, with standard errors estimated using maximum likelihood estimation (MLE).

\section{Simulation study}\label{s3:sim}

\subsection{Local versus spatial instruments}\label{ss3:localVsSpatialZ}

Our findings line up well with intuition from \cite{paciorek2010importance}, who finds that spatial confounding can be reduced, but only when the spatial treatment exists at a smaller spatial scale than the spatial confounder. That is, more local treatments can reduce more spatial confounding. This lines up quite well with the intuition developed below: that more local IVs with which to transform the treatment (using 2-stage least squares) provide more information with which to reduce confounding -- and therefore mitigate bias better than more spatially-dependent IVs.  

To illustrate this issue we conduct a simple simulation study under several different settings. We consider $\cal D$ to be the $30 \times 30$ grid on the unit square, and generate data as follows: 
\begin{singlespace}
\begin{align*}
    A_s &= \beta_1 Z_{local,s} + \beta_2 Z_{spatial,s} + \beta_3 U_s + \epsilon_{1,s} \\ \\
    Y_s &= \delta_1 A_s +  \delta_2 U_s + \delta_3 V_s + \epsilon_{2,s}
\end{align*}
\end{singlespace}
\noindent where $Z_{local,s},\epsilon_{1,s},\epsilon_{2,s} \stackrel{iid}{\sim} \mbox{Normal}(0,1)$ and $Z_{spatial}$, $U$ and $V$ are independent Gaussian processes with mean zero, variance one, and isotropic with exponential covariance function with range parameter $\phi=0.2$. For each setting, the coefficients $\beta_1$ and $\beta_2$ are tuned to achieve a specific cor$(Z_s,A_s)$ and cor$(Z_s,U_s)$ for both $Z_s = Z_{local}$ and $Z_s = Z_{spatial}$. The strength of the missing confounder is controlled by $\beta_3$ which is held fixed at 0.5, and $\delta_1$ and $\delta_2$ which are held fixed at 1. We will vary $\delta_3$ between 1 (spatially correlated response residuals) and 0 (i.i.d.~response residuals).  Datasets generated in which cor$(Z_s,A_s)$ and cor$(Z_s,U_s)$ deviate from the target by more than 0.02 are discarded. Both $Z_{local}$ and $Z_{spatial}$ are included simultaneously so that they can each be tested on the same dataset.  Henceforth, ``cor$(Z_s,A_s)= c$'' implies that cor$(Z_{local},A_s) =$ cor$(Z_{spatial},A_s)= c$.

In each setting we examine a total of six different models. For each simulated data set we fit the models separately with the Local IV, Spatial IV, and No IV (simply a response regression of $Y$ onto $A$). We repeat this once using i.i.d.~errors on the stage-1 and stage-2 regressions, and once using spatial errors on the stage-1 and stage-2 regressions. (For the No IV scenario there is no stage-1 regression.) Spatial errors are estimated as a Gaussian process with exponential covariance. All simulations settings are repeated 500 times. Each simulation examines four different metrics: bias, mean squared error, and the coverage percentage of the 95\% confidence intervals, and the average correlation between the instrument and the stage-2 regression residuals. The correlation with the stage-2 residuals is included, because this should be close to zero for effective instruments. Large correlations could indicate a non-valid IV.

The first scenario shown in Table \ref{table:strongValid} investigates the situation of a reasonably strong IV (cor$(Z_s,A_s) = 0.7$) which is perfectly valid (cor$(Z_s,U_s) = 0$). Here $\beta_1 = 1$, $\beta_2=1.1$, $\beta_3 = 0.5$, and $\delta_1 = \delta_2 = \delta_3 = 1$. This simulation demonstrates that with a strong and valid IV, local IVs can reduce the bias more than spatially-dependent IVs, and give substantially lower MSE. This is true for both i.i.d.~and spatial error models.  Note that because $\delta_3 =1$, the errors are spatially correlated and thus the spatial error model is the correctly specified model. This agrees well with the intuition that local IVs contain more information than spatial IVs. The increased information reduces the bias, but also greatly reduces the variability -- giving an even greater reduction in MSE. Of particular note is the low coverage on the spatial IV, (misspecified) i.i.d.~errors model.

\begin{table}[h!]
\caption{\textbf{Strong, valid IV scenario, with spatially correlated response residuals.} Cor$(Z_s,A_s) = 0.7$, cor$(Z_s,U_s) = 0$, and $\delta_3 = 1$. Simulation includes the no instrumental variable model, the local instrument model, and the spatial instrument model; using both i.i.d.~and spatial stage-1 and stage-2 error terms.  The Bias, MSE, and mean correlation values are multiplied by 10. Standard errors are included as subscripts.}
\label{table:strongValid}
\centering
\begin{singlespace}
\begin{center}
\begin{tabular}{ c|c|ccccc } 
\begin{tabular}{@{}c@{}}Instrument\\model\end{tabular} & 
\begin{tabular}{@{}c@{}}Residual\\model\end{tabular} &  
Bias &  MSE &  \begin{tabular}{@{}c@{}}95\%\\Cov.\end{tabular} & 
\begin{tabular}{@{}c@{}}Mean\\cor$(Z, \epsilon_2)$\end{tabular} \\ \hline
No IV &  \multirow{3}{*}{\begin{tabular}{@{}c@{}}I.i.d.\\errors\end{tabular}} & 
1.58$_{0.04}$ & 0.34$_{0.01}$ & 14.7$_{1.6}$ &  --- \\ 
Local IV & & 0.03$_{0.02}$ & 0.02$_{0.00}$ & 99.8$_{0.2}$ & 0.00$_{0.00}$ \\
Spatial IV & & -0.09$_{0.07}$ & 0.28$_{0.02}$ & 57.1$_{2.2}$ & 0.00$_{0.00}$ \\
 \hline
No IV & \multirow{3}{*}{\begin{tabular}{@{}c@{}}Spatial\\errors\end{tabular}} &
0.77$_{0.01}$ & 0.07$_{0.00}$ & 38.7$_{2.2}$ & --- \\
Local IV & & 0.02$_{0.02}$ &0.02$_{0.00}$ & 98.2$_{0.6}$ & 0.13$_{0.01}$ \\
Spatial IV & & -0.06$_{0.04}$ &0.07$_{0.00}$ &99.2$_{0.4}$ & -0.02$_{0.03}$ \\
 \hline
\end{tabular}
\end{center}
\end{singlespace}
\end{table}

The second scenario shown in Table \ref{table:weakValid} investigates the situation of a somewhat weaker IV (cor$(Z_s,A_s) = 0.3$) which is perfectly valid (cor$(Z_s,U_s) = 0$). Here $\beta_1 = \beta_2= 0.15$, $\beta_3 = 0.5$, and $\delta_1 = \delta_2 = \delta_3 = 1$. As expected, this simulation has much larger bias, but tells much the same story as Table \ref{table:strongValid}:  the local IV's generally improve upon the spatial IV's. Again note that the misspecified i.i.d.~errors model with the spatial IV gives poor coverage.

\begin{table}[h!]
\caption{\textbf{Weak, valid IV scenario, with spatially correlated response residuals.} Cor$(Z_s,A_s) = 0.3$, cor$(Z_s,U_s) = 0$, and $\delta_3=1$. Simulation includes a no instrumental variable model, the local instrument model, and the spatial instrument model; using both i.i.d.~and spatial stage-1 and stage-2 error terms.  The Bias, MSE, and mean correlation values are multiplied by 10. Standard errors are included as subscripts.}
\label{table:weakValid}
\centering
\begin{singlespace}
\begin{center}
\begin{tabular}{ c|c|ccccc } 
\begin{tabular}{@{}c@{}}Instrument\\model\end{tabular} & 
\begin{tabular}{@{}c@{}}Residual\\model\end{tabular} &  
Bias &  MSE &  \begin{tabular}{@{}c@{}}95\%\\Cov.\end{tabular} 
& \begin{tabular}{@{}c@{}}Mean\\cor$(Z, \epsilon_2)$\end{tabular} \\ \hline
No IV &  \multirow{3}{*}{\begin{tabular}{@{}c@{}}I.i.d.\\errors\end{tabular}} & 
 16.52$_{0.13}$ &  28.15$_{0.44}$ &  0.0$_{0.0}$ &  --- \\
Local IV & & -0.33$_{0.14}$ &  0.93$_{0.05}$ &  99.8$_{0.02}$ & 0.00$_{0.00}$ \\
Spatial IV & & 0.71$_{0.52}$ &  13.79$_{1.04}$ &  54.8$_{2.2}$ & 0.00$_{0.00}$ \\
 \hline
No IV & \multirow{3}{*}{\begin{tabular}{@{}c@{}}Spatial\\errors\end{tabular}} &
 13.62$_{0.06}$ &  18.71$_{0.16}$ &  0.0$_{0.0}$  &  --- \\
Local IV & & -0.08$_{0.12}$ &  0.74$_{0.04}$ &  96.4$_{0.8}$ &   -0.03$_{0.01}$ \\
Spatial IV & &  0.17$_{0.27}$ &  3.73$_{0.29}$ &  96.2$_{0.8}$ & 0.02$_{0.04}$ \\
 \hline
\end{tabular}
\end{center}
\end{singlespace}
\end{table}

We repeat this pair of simulations with i.i.d.~response errors in Tables \ref{table:strongValidD3eq0} and \ref{table:weakValidD3eq0}. The specification is identical to the previous two (i.e., the same values of $\beta_1$, $\beta_2$, $\delta_1$, and $\delta_2$) except that now $\delta_3$ is set to 0 giving independent errors.  The i.i.d.~response error setting tends to produce more similar bias levels between the local/spatial IV's.  Moreover, all IV methods seem to exhibit over-coverage.  However, the recurring theme remains:  the local IV's tend to improve upon the spatial IV's. Also of note, the misspecified spatial error models do not seem to produce low coverage (as the misspecified i.i.d.~error models did), making them a preferable choice if model misspecification is a concern. 

\begin{table}[h!]
\caption{\textbf{Strong, valid IV scenario, with i.i.d.~response residuals.} Cor$(Z_s,A_s) = 0.7$, cor$(Z_s,U_s) = 0$, and $\delta_3 = 1$.  Simulation includes a no instrumental variable model, the local instrument model, and the spatial instrument model; using both i.i.d.~and spatial stage-1 and stage-2 error terms.  The Bias, MSE, and mean correlation values are multiplied by 10. Standard errors are included as subscripts.}
\label{table:strongValidD3eq0}
\centering
\begin{singlespace}
\begin{center}
\begin{tabular}{ c|c|ccccc } 
\begin{tabular}{@{}c@{}}Instrument\\model\end{tabular} & 
\begin{tabular}{@{}c@{}}Residual\\model\end{tabular} &  
Bias &  MSE &  \begin{tabular}{@{}c@{}}95\%\\Cov.\end{tabular} 
& \begin{tabular}{@{}c@{}}Mean\\cor$(Z, \epsilon_2)$\end{tabular} \\ \hline
No IV &  \multirow{3}{*}{\begin{tabular}{@{}c@{}}I.i.d.\\errors\end{tabular}} & 
1.59$_{0.02}$ & 0.26$_{0.01}$ &  0.2$_{0.2}$ &  --- \\
Local IV && -0.02$_{0.02}$ &  0.01$_{0.00}$ &  100.0$_{0.0}$ & 0.00$_{0.00}$ \\
Spatial IV && -0.02$_{0.01}$ &  0.01$_{0.00}$ &  100.0$_{0.0}$ & 0.00$_{0.00}$ \\
 \hline
No IV & \multirow{3}{*}{\begin{tabular}{@{}c@{}}Spatial\\errors\end{tabular}} &
 0.86$_{0.01}$ &  0.08$_{0.00}$ &  18.8$_{1.7}$  &  --- \\
Local IV &&   0.01$_{0.02}$ &  0.01$_{0.00}$ &  98.0$_{0.6}$ &    0.12$_{0.01}$ \\
Spatial IV &&  -0.03$_{0.03}$ &  0.04$_{0.00}$ &  99.6$_{0.3}$ & 0.00$_{0.02}$ \\
 \hline
\end{tabular}
\end{center}
\end{singlespace}
\end{table}

\begin{table}[h!]
\caption{\textbf{Weak, valid IV scenario, with i.i.d.~response residuals.} Cor$(Z_s,A_s) = 0.3$, cor$(Z_s,U_s) = 0$, and $\delta_3=0$. Simulation includes a no instrumental variable model, the local instrument model, and the spatial instrument model; using both i.i.d.~and spatial stage-1 and stage-2 error terms.  The Bias, MSE, and mean correlation values are multiplied by 10. Standard errors are included as subscripts.}
\label{table:weakValidD3eq0}
\centering
\begin{singlespace}
\begin{center}
\begin{tabular}{ c|c|ccccc } 
\begin{tabular}{@{}c@{}}Instrument\\model\end{tabular} & 
\begin{tabular}{@{}c@{}}Residual\\model\end{tabular} &  
Bias &  MSE &  \begin{tabular}{@{}c@{}}95\%\\Cov.\end{tabular} 
& \begin{tabular}{@{}c@{}}Mean\\cor$(Z, \epsilon_2)$\end{tabular} \\ \hline
No IV &  \multirow{3}{*}{\begin{tabular}{@{}c@{}}I.i.d.\\errors\end{tabular}} & 
16.42$_{0.03}$ & 27.01$_{0.11}$ & 0.0$_{0.0}$ &    --- \\
Local IV & & -0.23$_{0.10}$ & 0.52$_{0.03}$ & 100.0$_{0.0}$ & 0.00$_{0.00}$ \\
Spatial IV & &0.20$_{0.11}$ & 0.64$_{0.04}$ & 99.2$_{0.4}$ & 0.00$_{0.00}$ \\
 \hline
No IV & \multirow{3}{*}{\begin{tabular}{@{}c@{}}Spatial\\errors\end{tabular}} &
 15.42$_{0.04}$ & 23.87$_{0.13}$ & 0.0$_{0.0}$ &     --- \\
Local IV & & -0.12$_{0.11}$ & 0.57$_{0.03}$ & 97.6$_{0.7}$&  -0.01$_{0.01}$ \\
Spatial IV & & -0.13$_{0.24}$ & 2.95$_{0.21}$ & 97.2$_{0.7}$ & 0.01$_{0.02}$ \\
 \hline
\end{tabular}
\end{center}
\end{singlespace}
\end{table}

\begin{figure}[h]
    \centering
    \includegraphics[width=1\textwidth]{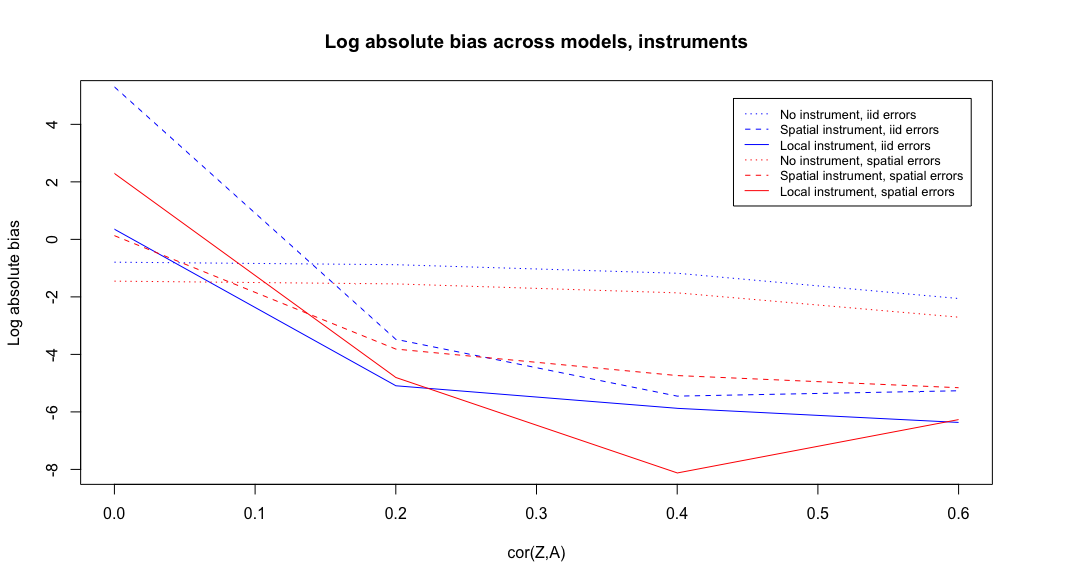}
    \caption{\textbf{Log absolute bias across models and instruments.} Cor$(Z_s,A_s)$ is varied between 0 and 0.6, with cor$(Z_s,U_s) = 0$. Blue lines represent models with i.i.d.~errors; red lines represent models with spatial errors. Dotted lines are models without an instrument; dashed lines are models using the spatial instrument; solid lines are models using the local instrument.}
    \label{fig:biasLines}
\end{figure}

Figure \ref{fig:biasLines} illustrates these trends over a range of cor$(Z_s,A_s)$ for valid $Z$ (cor$(Z_s,U_s)=0$) with the log absolute bias, $\log(|\hat{\delta} - \delta|)$. Intuitively the models with no instruments are relatively constant across cor$(Z_s,A_s)$. Also unsurprising is that the instrument models tend to improve as cor$(Z_s,A_s)$ increases. Moreover, for strong IV's (high cor$(Z_s,A_s)$), the local instruments outperform the spatial instruments in both the i.i.d.~and spatial error models. What is interesting is that for very weak IV's (low cor$(Z_s,A_s)$) the spatial IV, spatial error model out performs the local IV, spatial error model. For very weak IVs the non-instrument models universally outperform the instrument models. 

\subsection{Sensitivity Analysis} \label{ss3:squareSim}

Lastly, in the vein of a sensitivity analysis, we examine the situation in which the IV is not completely valid due to a non-zero correlation between $Z$ and $U$: 
\begin{singlespace}
\begin{align*}
      U  &\sim\gamma_1Z_{spatial}+\gamma_2Z_{local}+W\\ \\
      A      &\sim \beta_1 Z_{local} +\beta_2 Z_{spatial} +\beta_3 U +\epsilon_1\\ \\
      Y      &\sim \delta_1 A +  \delta_2 U + \delta_3 V + \epsilon_2
\end{align*}
\end{singlespace}
\noindent where $Z_{local,s},\epsilon_{1,s},\epsilon_{2,s} \stackrel{iid}{\sim}$ Normal$(0,1)$, and $Z_{spatial}$, $V$, and $W$ are independent Gaussian processes with mean zero, variance one, and isotropic with exponential covariance function with range parameter $\phi = 0.2$. This deviates from the previous specification only in that it creates dependence between both $Z$'s and $U$.  Again we tune coefficients ($\gamma_1 = \gamma_2 = 0.1$, $\beta_1 = \beta_2 = \delta_1 = \delta_2 = \delta_3 = 1$) and discard non-conforming datasets such that cor$(Z_s,A_s)=0.7$ and cor$(Z_s,U_s)=0.1$.  This corresponds to the situation of a strong IV which is slightly invalid due to its dependence with $U$.

\begin{table}[h!]
\caption{\textbf{Strong, invalid IV scenario, with spatially correlated response residuals.} Cor$(Z_s,A_s) = 0.7$, cor$(Z_s,U_s) = 0.1$. Simulation includes a no instrumental variable model, the local instrument model, and the spatial instrument model; using both i.i.d.~and spatial stage-1 and stage-2 error terms.  The Bias, MSE, and mean correlation values are multiplied by 10. Standard errors are included as subscripts. Note that, because $Z$ and $U$ are correlated here, $Z$ is \textit{not} a valid instrument, and is only included here as a sensitivity analysis.}
\label{table:strongInvalidIV}
\centering
\begin{singlespace}
\begin{center}
\begin{tabular}{ c|c|ccccc } 
\begin{tabular}{@{}c@{}}Instrument\\model\end{tabular} & 
\begin{tabular}{@{}c@{}}Residual\\model\end{tabular} &  
Bias &  MSE &  \begin{tabular}{@{}c@{}}95\%\\Cov.\end{tabular} 
& \begin{tabular}{@{}c@{}}Mean\\cor$(Z, \epsilon_2)$\end{tabular} \\ \hline
No instrument & \multirow{3}{*}{\begin{tabular}{@{}c@{}}I.i.d.\\errors\end{tabular}} & 
2.33$_{0.04}$ &0.61$_{0.02}$ & 2.8$_{0.8}$ & --- \\
Local instrument & & 0.81$_{0.02}$ &0.08$_{0.00}$ &89.7$_{1.4}$ & 0.00$_{0.00}$ \\
Spatial instrument & & 0.90$_{0.07}$ &0.28$_{0.02}$ &53.7$_{2.3}$ & 0.00$_{0.00}$ \\
 \hline
No instrument & \multirow{3}{*}{\begin{tabular}{@{}c@{}}Spatial\\errors\end{tabular}} &
1.57$_{0.01}$ &0.26$_{0.00}$ &0.0$_{0.0}$ & --- \\
Local instrument & & 0.90$_{0.02}$ &0.10$_{0.00}$ &45.6$_{2.3}$ & 0.05$_{0.01}$ \\
Spatial instrument & & 0.86$_{0.03}$ &0.13$_{0.01}$ &93.2$_{1.2}$ & 0.01$_{0.03}$ \\
 \hline
\end{tabular}
\end{center}
\end{singlespace}
\end{table}

Table \ref{table:strongInvalidIV} shows that with a strong but slightly invalid IV, the relative effectiveness between the local and spatial IVs becomes less clear, although they both reduce bias compared to the no-instrument method.  For the independent error models, the local instrument shows slightly less bias than the spatial error modes; for spatial error models the opposite is true. However, under both i.i.d.~and spatial error models, the MSE is reduced with the local instrument. Again, this illustrates that the local instrument reduces the variance of the estimator when compared to the spatial instrument, even if it does not always reduce the bias relative to the spatial instrument. 

Appendix A continues in this direction by exploring performance over a range of both IV strength (cor$(Z_s,A_s)$) as well as IV validity (cor$(Z_s,U_s)$).  Figure \ref{fig:logAbsBias} examines the log absolute bias across IV strength and validity.  Figure \ref{fig:logRelAbsBias} condenses this to show the improvement in log bias with the no-instrument model as reference. Lastly, Figure \ref{fig:coverageLess95} shows the coverage of the models over the IV strength/validity spectra. 

The intuition that we glean from these simulations is simply that the local instrument contains more information than the spatial instrument. When the information is sound (i.e., the instrument is valid) the local instrument reduces bias and variance. However, a consequence of this is that if the instrument used is \emph{not} valid (i.e., $Z$ is correlated with the confounder $U$), then the local instrument can do worse than its spatial counterpart. Therefore when the information is unsound (i.e., with an invalid instrument), the spatial instrument may actually perform better, as might non-instrument models for substantially invalid instruments.  

Additionally, the spatial error models appear to be a conservative choice over the i.i.d.~error models. Both models give appropriate coverage when they are the correctly specified model; however, the i.i.d.~error model gives very low coverage when misspecified (i.e., the true model has spatial errors) and the IV is spatially dependent. In other situations, however, the coverage between the i.i.d.~and spatial error models is very similar, despite the i.i.d.~error models giving smaller standard errors. When in doubt, the spatial error model should be used, and caution warranted when using the i.i.d.~error model.

\subsection{Simulation based on PM$_{2.5}$, cardiovascular mortality data} \label{ss3:realInspiredSim}

Before implementing this method on the observed PM$_{2.5}$ and cardiovascular mortality data, we test the performance on simulated data with similar spatial ranges and correlation structure. In particular, the observed IV and responses have moderate spatial correlation (practical ranges of 5.2 and 6.3 degrees respectively) while the observed treatment has substantial spatial dependence (practical range of 164 degrees). We use data generated on a $20 \times 20$ grid on the unit square, but maintain the spatial dependence of the observed $Z$, $A$ and $Y$ by requiring that, for each variable, the ratio of its practical range to the maximum distance between points on the grid is within 0.02 of the analogous observed-data statistic.  Similarly, the expected correlations between the variables are kept to within 0.035 of the observed correlations. The confounder $U$ is uncorrelated with $Z$ and given a spatial practical range roughly three times larger than the less spatial $Z$ and $Y$, but far smaller than the highly spatial $A$.
 
This is accomplished in several steps. $Z$, $U$, and $W$ are generated as mean-zero, zero-nugget, variance-one Gaussian processes that have exponential spatial covariance with range parameters $\phi =$ 0.0436, 0.15, and 1.8 respectively. $A$ is then set to $\sqrt{a}\cdot Z + \sqrt{b}\cdot U + \sqrt{c}\cdot W$, where $a=0.27, b=0.25, c=0.48$. $Y$ is set to $\sqrt{f}\cdot A + \sqrt{g}\cdot U + \sqrt{j}\cdot\epsilon$, where $f =0.063$, $g =0.023$, $j =0.914$, and $\epsilon$ has a standard normal distribution. The simulation includes 1,000 repetitions. As seen in Table \ref{table:realInspiredSim}, the IV methods perform quite well in this scenario. Bias is substantially reduced from the no IV models, and coverage is very good.

\begin{table}[h]
\caption{\textbf{Simulations with data generating mechanisms inspired by the real-life data application.} Standard errors are included as subscripts. The bias and MSE estimates and standard errors are multiplied by 100.}
\label{table:realInspiredSim}
\small
\centering
\begin{singlespace}
\begin{center}
\begin{tabular}{ r|cccc } 
model &  Bias &  MSE &  95\% Cov. \\ \hline
No IV, i.i.d.~errors    &  $11.08_{0.20}$ & $1.62_{0.05}$ & $56.1_{1.57}$ \\
With IV, i.i.d.~errors  &  $-0.10_{0.31}$ & $0.95_{0.04}$ & $95.0_{0.69}$ \\
\hline
No IV, spatial errors   &  $10.94_{0.20}$ & $1.58_{0.05}$ & $59.1_{1.55}$ \\
With IV, spatial errors &  $0.02_{0.32}$ & $1.01_{0.05}$ & $94.0_{0.75}$ \\
\hline
\end{tabular}
\end{center}
\end{singlespace}
\end{table}

\section{Assessing the causal impact of air pollution on cardiovascular mortality} \label{s3:realData}

\subsection{Data}

We use the proposed method to estimate the average effect of the time trend of air pollution (estimated log PM$_{2.5}$) on the time trend of the cardiovascular mortality rate (CMR) at the county level over 1990-2010. The instrument used is the log of the local primary PM$_{2.5}$ emissions. We theorize that this is likely a valid instrument because it will primarily be a function of local factors related to traffic, and will be largely independent of the macro spatial confounders affecting total PM$_{2.5}$ and CMR.  Specifically, we theorize that local primary PM$_{2.5}$ emissions are only related to CMR through their effect on total PM$_{2.5}$, and not through other confounders that we have not accounted for. This hypothesis notwithstanding, completely valid instrumental variables are difficult to find, and we acknowledge that this instrument may be imperfect. 

The CMR data are annual averages of the age-adjusted rates of death resulting from cardiovascular issues per 100,000 people. Because of privacy concerns, only the 2,132 counties with a population of at least 20,000 are included. The unlogged PM$_{2.5}$ treatment variable measures the annual average of the amount of airborne particulate matter smaller than 2.5 $\mu$m, as measured in units of $\mu g/m^3$. The data are estimates from the Community Multiscale Air Quality (CMAQ) model implemented by the Environmental Protection Agency, evaluated on the grid and averaged across each county \citep{us_epa_office_of_research_and_developmen_2020_4081737}. The cardiovascular mortality rate and PM$_{2.5}$ levels are both taken from \cite{wyatt2020annual}. The instrumental variable used is the time trend in local primary PM$_{2.5}$ emissions. This encapsulates all PM$_{2.5}$ emissions (including from cars, power plants, wildfires, etc.) and is measured in average grams per day. \citep{peterson2020impact}.

In addition to the three main variables, \cite{wyatt2020annual} provides a number of county-level covariates collected for US Census years 1990, 2000, and 2010, including percentage of households below poverty line, median household income, percentage of individuals 25 years and over with a high school education, civilian unemployment rate, percent female households with no spouse, percentage vacant houses, and percent owner occupied housing.

\begin{figure}
    \centering
    \makebox[\textwidth][c]{\includegraphics[width=.95\textwidth]{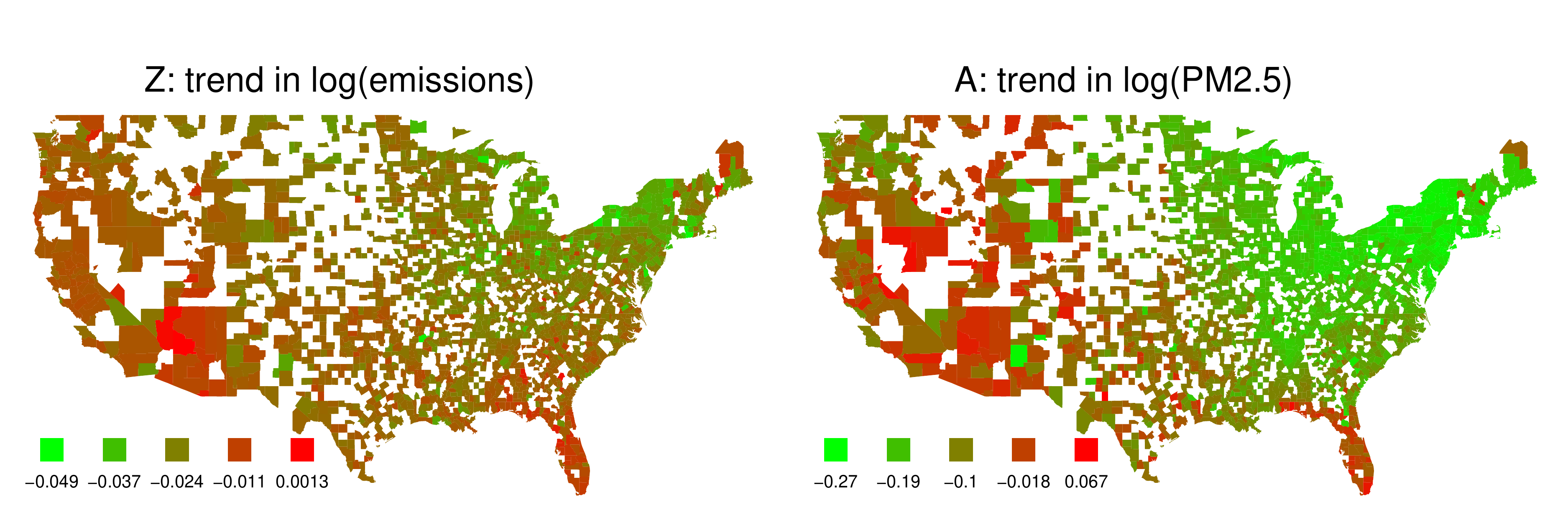}}
    \makebox[\textwidth][c]{\includegraphics[width=.95\textwidth]{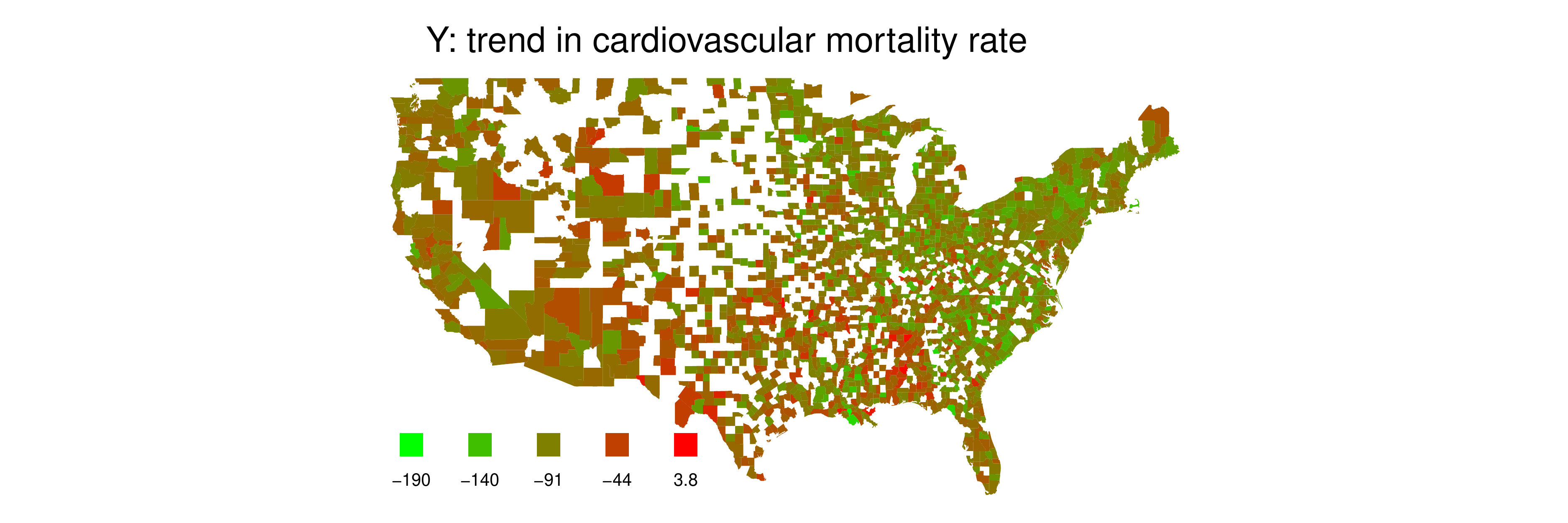}}
    
    \vspace{-5mm}
    \caption{\textbf{Increase per decade in three main variables.} 
    Unlogged local emissions ($Z$) are given in average grams/day. 
    Unlogged PM$_{2.5}$ ($A$) is given in $\mu g/m^3$.
    Cardiovascular mortality rate ($Y$) is given as the number of individuals per 100,000 people.}
    \label{fig:dataSnapshot}
\end{figure}

All variables are transformed into their slopes or trends per decade at each location before the method is implemented. That is, at each location, $Z$, $A$, $Y$, and the covariates $X_j$ are each regressed on time as measured by 10-year units ($t = (\text{year} - 1990)/10$), and the resulting trend coefficient is used as the datapoint in the model. An illustration of the decadal slope data for the three main variables is given in Figure \ref{fig:dataSnapshot}. 

Table \ref{table:spatialPropertiesZAY} gives the estimated spatial properties for these three variables. In estimating these parameters, the variables are assumed to be Gaussian processes with exponential spatial covariance. The percentage of error attributed to spatial variability is calculated as $\sigma^2/(\tau^2 + \sigma^2)$ where $\tau^2$ is the estimated nugget parameter and $\sigma^2$ is the estimated partial sill parameter. $\phi$ is the estimated range parameter, in one-degree units. Practical range is the estimated distance at which correlation has decayed to 0.05, and is also given in units of one degree.

\begin{table}[h!]
\caption{\textbf{Estimated spatial properties of observed instrument ($Z$), treatment ($A$), and response ($Y$).}
Practical range is the distance at which correlation has decayed to $0.05$. Both  $\phi$  and practical range are given in degrees of latitude/longitude.
}
\label{table:spatialPropertiesZAY}
\centering
\begin{singlespace}
\begin{center}
\begin{tabular}{ c|ccc } 
Variable & \begin{tabular}{@{}c@{}}\% of error attributed\\to spatial variability\end{tabular}  &  $\phi$ & Practical range \\ \hline
Log primary emissions ($Z$) & 100\%  & 1.747 & 5.234 \\
Log total PM$_{2.5}$ ($A$)       & 100\%    & 54.70 & 163.9 \\
CMR ($Y$)                   & 30.6\%  & 2.113 & 6.330 \\
\hline
\end{tabular}
\end{center}
\end{singlespace}
\end{table}

The only non-slope variables used are the $A$-intercept from these location-specific regressions (which is included as a covariate in the first stage), and the $Y$-intercept from these location-specific regressions (which is used as a covariate in the second stage). These are included to provide information about the original levels of PM$_{2.5}$ and cardiovascular mortality at each location.

\subsection{Results}\label{ss3:results}

Table \ref{realTable} gives the estimates from our four models on the PM$_{2.5}$ and mortality data. Of the four models runs, only the No IV, i.i.d.~error model has a statistically significant coefficient for $\delta_1$. This coefficient is positive, which implies a decrease in air pollution is associated with a decrease in CMR. When spatial errors are incorporated on the No IV model, the confidence interval widens  and the estimated coefficient is not statistically significant. However the two confidence intervals are largely overlapping. The With IV, spatial error model has very large standard errors and the estimated coefficient is not significant.  The  With-IV, i.i.d.\ error model has smaller standard errors, but is still not significant. Of particular note in the With IV, spatial error model is the non-zero correlation between the observed IV and stage-2 residuals. While IV models that include a spatial error do not require this correlation to be zero (as i.i.d.~error models do), this is possibly an indication of a poor stage-1 model fit or an invalid instrument. Additionally, a difficulty with the real data example is the inability to test for the unmeasured confounding. The scenario where there is no spatial confounding was not represented in our simulations.

The simulation study showed that for data with spatially-correlated treatment and responses, the non-spatial model can give low coverage when misspecified; whereras the spatial error IV models give reliable coverage. The simulations also show that we can get more conservative but reliable results using a local IV. In the analysis of PM$_{2.5}$ and CVM,  the models with local IV variables suggest a positive effect, but the confidence intervals tend to be conservative, and possibly as a result these intervals include zero.

\begin{table}
\caption{\textbf{Estimated causal effect of increase in PM$_{2.5}$ trend on increase in CMR trend.} Here $\eps_{S2}$ refers to the residuals from the stage-2 regressions. Similarly, AIC(S1) and AIC(S2) refer to the AIC from the stage-1 and stage-2 regressions.}
\label{realTable}
\footnotesize
\centering
\begin{tabular}{r|ccccc}
\hline
model & $\boldsymbol{\delta_1}$ CI & cor$(Z,\eps_{S2})$ & AIC(S1) & AIC(S2) \\
\hline
No IV, i.i.d.~errors & 20.14 (5.22, 35.06) & 0.00 &  --- & 18,176 \\ 
No IV, spatial errors & -10.39 (-45.47, 24.69) & 0.13 &  --- & 17,589 \\ 
With IV, i.i.d.~errors & 28.44 (-8.29, 65.16) & 0.00 & -8,934&18,181\\  
With IV, spatial errors & 135.32 (-246.34, 516.98) & 0.09 & -14,974&17,588\\ 
\hline
\end{tabular}
\end{table}

\section{Extending to the interference setting}\label{s3:introInterference}

We briefly extend our method to the setting of spatial interference -- that is, when the local treatment can affect nearby responses. For brevity, we don't apply this method to the air pollution/CMR dataset. 

\subsection{Potential outcomes, interference, and instrumental variables revisited}\label{ss3:prelimsInterference}

We now consider the same variables $A_s$, $Y_s$, and $Z_s$ observed at the $n$ points $\mathcal{S} = \{s_1, \ldots, s_n\} \subset \mathcal{D} \subset \mathbb{R}^2$ as in Section \ref{s3:prelims}, but in this section we will allow potential \emph{spatial interference} \citep{giffin2020generalized, reich2020review}.  That is, nearby treatments may affect the response at location $s$.  This invalidates Assumption \ref{a:noInterference}, and thus requires new methods and assumptions.

The fundamental difficulty with interference is that each distinct set of individual treatments $a = \{a_1,\ldots,a_n\}\in \mathbb{R}^n$ represents a distinct treatment which must be considered. Using the potential outcomes framework, even for the simplified case of binary $a_s$, there would be $2^n$ potential outcomes $Y_s(a)$ to consider. To address this, we make a simplifying assumption on the form of interference that the treatment can only affect $Y_s$ through the following two mechanisms:

\begin{align}
    \text{Direct effect:}~~~a_s, \qquad \qquad ~~~
    \text{Indirect effect:}~~~\tilde{a}_s = \sum_{s' \ne s }
    \omega \left( \| s-s' \| \right)
    a_{s'},
\end{align}
where $\omega(d)$ is a kernel function of distance, with weights that sum to 1 for each $s$. For the remainder of the paper, we use the $\omega(d) = d^{-1} \1(d < T)$ for maximum distance $T$, with weights then standardized to sum to one. With this simplification, the potential outcome can be rewritten as $Y_s(a_s, \tilde{a}_s)$. 

We now replace Assumption \ref{a:noInterference} with an assumption on the form of interference using our two treatment effect mechanisms. 

\begin{assumption}[Interference form] \label{a:nterferenceForm}
The potential outcome $Y_s(a)$ depends only on $a_s$ and $\tilde{a}_s$.  That is, $Y_s(a) = Y_s(a_s, \tilde{a}_s)$.
\end{assumption}

\begin{assumption}[Marginal Structural Model]
The potential outcomes for $A(z)$ and $Y(a)$ take the following form:

\begin{align}
  Y_s(a_s, \tilde{a}_s) &= \beta_0 + \delta_1 a_s + \delta_2 \tilde{a}_s + U_s + \epsilon_{1,s}, \label{eq:Yaform}\\
    A_s(z) &= \gamma_0 + \delta_3 z_s + \gamma_1 U_s + \epsilon_{2,s},\label{eq:Azform}
\end{align}
\noindent where $U$ is an unobserved, potentially spatially-dependent confounder with mean zero, and $\epsilon_1$ and $\epsilon_2$ are mean zero error processes. Both error terms are potentially spatially dependent.
\end{assumption}

Moreover, we update our consistency assumption:

\begin{assumption}[Consistency] 
The potential outcome $Y_s(a,\tilde{a})$ is the equal to the response $Y_s$ that would be observed under treatments $A_s$ and $\tilde{A}_s$.
\end{assumption}

\subsection{Identification}\label{ss3:identificationInterference}
If we can establish that 
$\hat{A}_s = \E [A_s | Z]$ and $\hat{\tilde{A}}_s = \E [\tilde{A}_s | Z]$, 
we can take the conditional expectation of (\ref{eq:Yaform}), substituting $A_s$ and $\tilde{A}_s$ for $a_s$ and $\tilde{a}_s$, and write

\begin{align*}
\E [ Y_s(A_s, \tilde{A}) ~|~ Z] 
&= \E [ \beta_0 + \delta_1 A_s + \delta_2 \tilde{A}_s + U_s + 
\epsilon_{1,s} ~|~ Z ] \\
&= \beta_0 + \delta_1 \E [A_s|Z] + \delta_2 \E [ \tilde{A}_s | Z] + 
\E [U_s|Z] + \E[\epsilon_{1,s} | Z] \\
&=  \beta_0 + \delta_1 \E [A_s|Z] + \delta_2 \E [\tilde{A}_s|Z] \\
&=  \beta_0 + \delta_1 \hat{A}_s(Z)  + \delta_2 \hat{\tilde{A}}_s(Z).
\end{align*}

The first equality follows from (\ref{eq:Yaform}), and the second from Assumption \ref{a:validIV}. From this we see that a regression from the observed $Y_s$ onto $\hat{A}_s$ and $\hat{\tilde{A}}_s$ still recovers unbiased estimates of the causal parameter values $\delta_1$ and $\delta_2$. Two methods to create $\hat{A}_s$ and $\hat{\tilde{A}}_s$ that meet the above requirement are discussed in \ref{ss3:2sls}.

\subsection{Instrumental variables for the interference effect} \label{ss3:2sls}

We have established in Section \ref{s3:sim} the efficacy of instruments in mitigating confounding in our direct treatment effect. This section aims to establish an appropriate instrument for the indirect treatment $\tilde{A}_s = \sum_{s' \ne s} \omega (\|s-s'\|)A_{s'}$. We will explore two methods of incorporating $Z$ into $\tilde{A}$. Our estimation procedure first involves computing $\hat{A}$ as in Section \ref{s3:prelims} and then $\hat{\tilde{A}}$ to satisfy $\hat{\tilde{A}}_s = \E [\tilde{A}_s | Z]$.  These first-stage estimates are then using in the second stage regression $$Y = \beta_0 + \delta_1 \hat{A} + \delta_2 \hat{\tilde{A}} + e_1.$$ Below we discuss two methods to estimates $\hat{\tilde{A}}_s$.

\subsubsection{Type-1 spillover: kernel over fitted treatments} \label{type1spillover}
The first spillover type is given by
\begin{align*}
    \hat{\tilde{A}}_s &= 
    \sum_{s' \ne s }
    \omega \left( \frac{\| s-s' \|}{\tau} \right) \hat{A}_{s'} \qquad
    \text{ where } \hat{A}_{s'}= \hat{\gamma}_0 + \hat{\delta}_3 Z_{s'}, 
\end{align*}
This takes place in three stages: First, $A_s$ is regressed onto $Z_s$, giving the fitted values $\hat{A}_s$ as in Section \ref{ss3:estimation}.  Then, at each location $s$, a kernel-weighted average of the nearby $\hat{A}$ values is taken to obtain $\hat{\tilde{A}}_s$. Finally, the response regression regresses $Y_s$ onto $\hat{A}_s$ and $\hat{\tilde{A}}_s$ for final causal estimates of the treatment effects.

\subsubsection{Type-2 spillover:~~separate 1$^{st}$-stage spillover regression} \label{type2spillover}
The second spillover type is given by
\begin{align*}
        \hat{\tilde{A}}_s &= \hat{\gamma}_1 + \hat{\delta}_4 \tilde{Z}_s, \\
    \tilde{Z}_s &= \sum_{s' \ne s } \omega\left(\frac{\|s-s'\|}{\tau}\right) Z_{s'}.
\end{align*}
This also takes place in 3 stages: First, a kernel-weighted average is taken over the $Z_s$'s to give $\tilde{Z}_s$. Then $\tilde{A}_s$ is regressed onto $\tilde{Z}_s$, and $A_s$ is regressed onto $Z_s$ to get $\hat{\tilde{A}}_s$ and $\hat{A}_s$.  Finally, the response regression regresses $Y_s$ onto $\hat{A}_s$ and $\hat{\tilde{A}}_s$ for final causal estimates of the treatment effects.

Because i.i.d.~and spatial error regressions are unbiased \citep{basu1994regression}, it is clear that both $\hat{\tilde{A}}$ satisfy
$\tilde{\hat{A}}_s = \E [ \tilde{A}_{s} | Z ].$

\section{Simulation study with interference}\label{s3:simInterference}

We use a simulation to assess the accuracy and coverage of the methods described in Section \ref{ss3:2sls}. Variables are generated on a $32\times 32$ grid of points on $[0,1] \times [0,1]$. The confounder $U$, spatial error $V$, and instrument $Z$ are generated as independent mean-zero, variance-one, Gaussian processes with  isotropic exponential covariance, and no nugget effect. $V$ has spatial range parameter $\phi_V = 0.1$; while for $U$ and $V$ all combinations of $\phi_Z = 0.05,0.2$ and $\phi_U = 0.05,0.2$ are explored, to examine the effects of spatial range on performance. ($\phi = 0.2$ can be thought of as moderate spatial dependence; $\phi = 0.05$ is minimal spatial dependence.) $A$ and $Y$ are generated with the linear models 
\begin{align}
Y_s &= \beta_0 + \delta_1 A_s + \delta_2 \tilde{A}_s + \beta_1 U_s + \beta_2 V_s + \epsilon_{1,s},\\
A_s &= \gamma_0 + \gamma_1 Z_s + \gamma_2 U_s + \epsilon_{2,s},
\end{align}
where $\epsilon_{1,s}$ and $\epsilon_{2,s}$ are i.i.d.~standard normal error. In all simulations, $\beta_0 = \gamma_0 = 0$,  $\delta_1 = \delta_2 =  \beta_2 = 1$, and $\beta_1 = \gamma_2 = 2$.  We vary $\gamma_1$ to assess the performance under difference scenarios as described below. Lastly, $\tilde{A}_s$ is the distance-weighted sum of the ``queen'' neighbors to $s$ -- that is, the directly adjacent neighbors including diagonals. These values are weighted by the reciprocal of their distance to $s$ before the weights are standardized to sum to one.

The models are run under both a weak instrument setting in which cor$(Z_s,A_s) = 0.25$, and a strong instrument setting in which this correlation is increased to 0.75. This is achieved by tuning $\gamma_1$ to $0.7$ and $2.3$ respectively, and discarding any datasets that give a correlation more than 0.02 from the target correlation. The simulation is repeated 1,000 times to assess accuracy and coverage.

\subsection{Competing methods and metrics}\label{s3:sim:methods}

Each simulation compares four models:
\begin{itemize}
    \item \textbf{No instrument:} $A$ and $\tilde{A}$ enter the response regression directly, with no stage-1 regression.
     \item \textbf{Type-0 spillover:} The response regression uses $\hat{A}$ from the stage-1 regression with naive $\tilde{A}$.
    \item \textbf{Type-1 spillover:} The response regression uses $\hat{A}$ and Type-1 spillover to calculate $\hat{\tilde{A}}$, as in Section \ref{type1spillover}.
    \item \textbf{Type-2 spillover:} The response regression uses $\hat{A}$ uses Type-2 spillover to calculate $\hat{\tilde{A}}$, as in Section \ref{type2spillover}.
\end{itemize}
In all models, both stage-1 and stage-2 regressions assume spatial errors.

\subsection{Results}\label{s3:sim:results}

\begin{table}
\caption{\textbf{Accuracy and coverage across simulation different simulation settings.} Bias values are multiplied by 10. The weak instrument setting has cor$(Z_s,A_s) = 0.25$; the strong instrument setting has has cor$(Z_s,A_s) = 0.75$. Here $\phi_U$ and $\phi_Z$ correspond to the spatial range parameters in the confounder $U$ and the instrument $Z$. Each setting includes 1,000 repetitions.}
\label{simsTable}
\scriptsize
\centering
\begin{tabular}{c|r|cccc|cccc}
&&
\multicolumn{4}{c}{\underline{Weak instrument}} & 
\multicolumn{4}{|c}{\underline{Strong instrument}} \\
&&
\multicolumn{2}{c}{\underline{$~~~~~~~~~~~~\delta_1~~~~~~~~~~~~$}} & 
\multicolumn{2}{c|}{\underline{$~~~~~~~~~~~~\delta_2~~~~~~~~~~~~$}} &
\multicolumn{2}{c}{\underline{$~~~~~~~~~~~~\delta_1~~~~~~~~~~~~$}} & 
\multicolumn{2}{c}{\underline{$~~~~~~~~~~~~\delta_2~~~~~~~~~~~~$}} \\
&Model & 
Bias & 95\% Cov. & Bias & 95\% Cov. &
Bias & 95\% Cov. & Bias & 95\% Cov. \\
\hline
& No Instrument & 3.84$_{0.01}$ & 0.0$_{0.0}$ & 4.92$_{0.02}$ & 0.0$_{0.00}$ & 2.50$_{0.01}$ & 0.0$_{0.0}$ & 1.83$_{0.02}$ & 15.7$_{1.15}$ \\
$\phi_U=0.2$,& Type-0 Spillover &-3.56$_{0.21}$ & 76.1$_{1.4}$ & -5.56$_{0.45}$ & 0.0$_{0.00}$ & -0.04$_{0.02}$ & 97.3$_{0.5}$ & -1.47$_{0.13}$ & 41.7$_{1.56}$ \\
$\phi_Z=0.2$& Type-1 Spillover &-0.15$_{0.06}$ & 99.9$_{0.1}$ & -0.94$_{0.21}$ & 96.2$_{0.60}$ & 0.00$_{0.02}$ & 99.4$_{0.2}$ & 0.15$_{0.06}$ & 96.0$_{0.62}$ \\
& Type-2 Spillover & -0.15$_{0.06}$ & 99.9$_{0.1}$ & 1.13$_{0.26}$ & 95.4$_{0.66}$ & 0.00$_{0.02}$ & 99.4$_{0.2}$ & -0.40$_{0.07}$ & 91.9$_{0.86}$ \\
\hline
& No Instrument & 6.60$_{0.01}$ & 0.0$_{0.0}$ & 2.82$_{0.02}$ & 0.6$_{0.24}$ & 5.16$_{0.01}$ & 0.0$_{0.0}$ & 0.21$_{0.02}$ & 90.2$_{0.94}$ \\
$\phi_U=0.2$,& Type-0 Spillover &-14.64$_{0.16}$ & 6.5$_{0.8}$ & 16.87$_{0.30}$ & 0.0$_{0.00}$ & -0.56$_{0.08}$ & 81.9$_{1.2}$ & -2.69$_{0.24}$ & 19.6$_{1.26}$ \\
$\phi_Z=0.2$& Type-1 Spillover &-0.20$_{0.08}$ & 99.7$_{0.2}$ & -0.72$_{0.25}$ & 98.2$_{0.42}$ & -0.02$_{0.02}$ & 99.4$_{0.2}$ & 0.59$_{0.07}$ & 95.6$_{0.65}$ \\
& Type-2 Spillover &  -0.20$_{0.08}$ & 99.7$_{0.2}$ & -0.22$_{0.26}$ & 98.6$_{0.37}$ & -0.02$_{0.02}$ & 99.4$_{0.2}$ & 0.20$_{0.06}$ & 98.1$_{0.43}$ \\
\hline
& No Instrument & 3.55$_{0.01}$ & 0.0$_{0.0}$ & 4.93$_{0.02}$ & 0.0$_{0.00}$ & 1.48$_{0.01}$ & 0.0$_{0.0}$ & 1.64$_{0.02}$ & 14.7$_{1.12}$ \\
$\phi_U=0.2$,& Type-0 Spillover &-1.37$_{0.06}$ & 89.4$_{1.0}$ & -9.48$_{0.28}$ & 0.0$_{0.00}$ & 0.01$_{0.01}$ & 99.8$_{0.1}$ & -0.46$_{0.05}$ & 71.2$_{1.43}$ \\
$\phi_Z=0.2$& Type-1 Spillover &-0.13$_{0.03}$ & 99.9$_{0.1}$ & -1.31$_{0.13}$ & 95.1$_{0.68}$ & 0.02$_{0.01}$ & 99.6$_{0.2}$ & 0.01$_{0.04}$ & 95.1$_{0.68}$ \\
& Type-2 Spillover &  -0.13$_{0.03}$ & 99.9$_{0.1}$ & 5.68$_{0.33}$ & 90.7$_{0.92}$ & 0.02$_{0.01}$ & 99.6$_{0.2}$ & -0.09$_{0.05}$ & 88.3$_{1.02}$ \\
\hline
& No Instrument & 6.29$_{0.01}$ & 0.0$_{0.0}$ & 2.88$_{0.02}$ & 0.2$_{0.14}$ & 3.49$_{0.01}$ & 0.0$_{0.0}$ & 0.67$_{0.02}$ & 74.6$_{1.38}$ \\
$\phi_U=0.2$,& Type-0 Spillover &-8.28$_{0.10}$ & 13.0$_{1.1}$ & 14.28$_{0.39}$ & 0.0$_{0.00}$ & 0.00$_{0.01}$ & 99.5$_{0.2}$ & -0.10$_{0.10}$ & 49.5$_{1.58}$ \\
$\phi_Z=0.2$& Type-1 Spillover &-0.47$_{0.04}$ & 99.8$_{0.1}$ & -2.19$_{0.19}$ & 96.5$_{0.58}$ & 0.09$_{0.01}$ & 99.8$_{0.1}$ & 0.60$_{0.05}$ & 95.0$_{0.69}$ \\
& Type-2 Spillover & -0.47$_{0.04}$ & 99.8$_{0.1}$ & 0.30$_{0.25}$ & 97.5$_{0.49}$ & 0.09$_{0.01}$ & 99.8$_{0.1}$ & 0.12$_{0.05}$ & 97.4$_{0.50}$ 
\end{tabular}
\end{table}

\normalsize

Table \ref{simsTable} shows the simulation results. Generally, the proposed Type-1 and Type-2 spillover methods compare well to the competitor models in terms of both bias and coverage. With few exceptions, the coverage for the proposed methods under different settings is very close to 95\%, although it is somewhat conservative in some cases. The exception to this is that the Type-2 Spillover model produces slightly anti-conservative coverage on $\delta_2$ when $Z$ has greater spatial dependence than $U$. Again with few exceptions, the bias for the proposed methods is considerably smaller for both $\delta_1$ and $\delta_2$. 
The Type-0 model often performs well for estimating the direct effect $\delta_1$, as this model uses the correct $\hat{A}$ direct effect. The No Instrument model performs poorly throughout. The situation in which the proposed models clearly beat out the other models are those where there is moderate spatial confounding in both $U$ and $Z$. As expected, all models perform better under the strong instrument form. (Paradoxically, this is also true of the No Instrument model. However, this is simply a result of the form of $A$ being less noisy when $Z$ is a more spatially-dependent process.) 

Naturally, the Type-1 and Type-2 Spillover models seem to give identical bias and coverage with regards to $\delta_1$; their differences emerge in the $\delta_2$ performance. While they both generally perform well, the $\delta_2$-coverage on the Type-2 Spillover appears to dip below 90\% when $Z$ is spatially correlated, making Type-1 Spillover the preferable of the two methods.

\section{Discussion}\label{s3:discussion}

This paper examines the efficacy of using instrumental variables in the context of spatial confounding. In particular, it compares instruments with different degrees of spatial correlation, and provides a sensitivity analysis for non-valid instruments.  The simulations largely support the notion that while both spatial and local instruments mitigate spatial confounding, spatial instruments provide less information.  Therefore, they tend to mitigate bias less than local instruments. 
The method is illustrated with an analysis of the effect of changes in air pollution (PM$_{2.5}$) on changes in cardiovascular mortality across the United States from 1990-2010. Finally, the method is extended to allow for spatial interference, by applying some additional machinery and assumptions. A simulation study details the continued effectiveness in this setting.

This analysis has several limitations: our simulations in Section \ref{s3:sim} validate the intuition that local IVs provide more information than spatially-dependent IV; however, a rigorous theoretical explanation of this phenomenon would provide more clarity. Additionally, as with most real data analyses using instrumental variables, the validity of the instrument is always a concern. The simulations in this manuscript included scenarios where the simulated data mimicked the spatial ranges and correlations between $Z$, $A$, and $Y$ in the data example. More theoretical work to justify the notion that local emissions are uncorrelated with the relevant confounders would add certainty to our intuition. 

One key direction for future work in this area will be to thoroughly investigate the relationship between the four relevant variables from this framework  (IV, treatment, confounder, and response) with regards to their spatial dependence. For example, this analysis looks at a spatial confounder and compares spatial versus non-spatial IVs. Many other non-IV analyses have focused on a spatial confounder, occasionally also addressing treatment spatial dependence. However, no analysis has yet comprehensively investigated all permutations of spatial/non-spatial versions of these variables. While ambitious, this type of broad analysis would be very helpful for tying together these disparate studies.

\bibliographystyle{apalike}
\bibliography{refs} 

\begin{thebibliography}{}

\bibitem[Angrist et~al., 1996]{angrist1996identification}
Angrist, J.~D., Imbens, G.~W., and Rubin, D.~B. (1996).
\newblock Identification of causal effects using instrumental variables.
\newblock {\em Journal of the American Statistical Association},
  91(434):444--455.

\bibitem[Basu and Reinsel, 1994]{basu1994regression}
Basu, S. and Reinsel, G.~C. (1994).
\newblock Regression models with spatially correlated errors.
\newblock {\em Journal of the American Statistical Association},
  89(425):88--99.

\bibitem[Betz et~al., 2020]{betz2017spatial}
Betz, T., Cook, S.~J., and Hollenbach, F.~M. (2020).
\newblock Spatial interdependence and instrumental variable models.
\newblock {\em Political Science Research and Methods}, 8(4):646–661.

\bibitem[Davis et~al., 2019]{davis2019addressing}
Davis, M.~L., Neelon, B., Nietert, P.~J., Hunt, K.~J., Burgette, L.~F., Lawson,
  A.~B., and Egede, L.~E. (2019).
\newblock Addressing geographic confounding through spatial propensity scores:
  a study of racial disparities in diabetes.
\newblock {\em Statistical Methods in Medical Research}, 28(3):734--748.

\bibitem[Giffin et~al., 2020]{giffin2020generalized}
Giffin, A., Reich, B., Yang, S., and Rappold, A. (2020).
\newblock Generalized propensity score approach to causal inference with
  spatial interference.
\newblock {\em arXiv preprint arXiv:2007.00106}.

\bibitem[Hodges and Reich, 2010]{hodges2010adding}
Hodges, J.~S. and Reich, B.~J. (2010).
\newblock Adding spatially-correlated errors can mess up the fixed effect you
  love.
\newblock {\em The American Statistician}, 64(4):325--334.

\bibitem[Jarner et~al., 2002]{jarner2002estimation}
Jarner, M.~F., Diggle, P., and Chetwynd, A.~G. (2002).
\newblock Estimation of spatial variation in risk using matched case-control
  data.
\newblock {\em Biometrical Journal: Journal of Mathematical Methods in
  Biosciences}, 44(8):936--945.

\bibitem[Kelejian and Piras, 2014]{kelejian2014estimation}
Kelejian, H.~H. and Piras, G. (2014).
\newblock Estimation of spatial models with endogenous weighting matrices, and
  an application to a demand model for cigarettes.
\newblock {\em Regional Science and Urban Economics}, 46:140--149.

\bibitem[Kelejian and Prucha, 1998]{kelejian1998generalized}
Kelejian, H.~H. and Prucha, I.~R. (1998).
\newblock A generalized spatial two-stage least squares procedure for
  estimating a spatial autoregressive model with autoregressive disturbances.
\newblock {\em The Journal of Real Estate Finance and Economics},
  17(1):99--121.

\bibitem[Kelejian et~al., 2004]{kelejian2004instrumental}
Kelejian, H.~H., Prucha, I.~R., and Yuzefovich, Y. (2004).
\newblock Instrumental variable estimation of a spatial autoregressive model
  with autoregressive disturbances: Large and small sample results.
\newblock {\em Advances in Econometrics: Spatial and Spatio-Temporal
  econometrics}, pages 163--198.

\bibitem[Keller and Szpiro, 2020]{keller2020selecting}
Keller, J.~P. and Szpiro, A.~A. (2020).
\newblock Selecting a scale for spatial confounding adjustment.
\newblock {\em Journal of the Royal Statistical Society, Statistics in Society,
  Series A}, 183(Part 3):1121--1143.

\bibitem[Lee, 2003]{lee2003best}
Lee, L.-f. (2003).
\newblock Best spatial two-stage least squares estimators for a spatial
  autoregressive model with autoregressive disturbances.
\newblock {\em Econometric Reviews}, 22(4):307--335.

\bibitem[Paciorek, 2010]{paciorek2010importance}
Paciorek, C.~J. (2010).
\newblock The importance of scale for spatial-confounding bias and precision of
  spatial regression estimators.
\newblock {\em Statistical science: a review journal of the Institute of
  Mathematical Statistics}, 25(1):107.

\bibitem[Papadogeorgou et~al., 2019]{papadogeorgou2019adjusting}
Papadogeorgou, G., Choirat, C., and Zigler, C.~M. (2019).
\newblock Adjusting for unmeasured spatial confounding with distance adjusted
  propensity score matching.
\newblock {\em Biostatistics}, 20(2):256--272.

\bibitem[Peterson et~al., 2020]{peterson2020impact}
Peterson, G. C.~L., Hogrefe, C., Corrigan, A.~E., Neas, L.~M., Mathur, R., and
  Rappold, A.~G. (2020).
\newblock Impact of reductions in emissions from major source sectors on fine
  particulate matter--related cardiovascular mortality.
\newblock {\em Environmental health perspectives}, 128(1):017005.

\bibitem[Piras et~al., 2010]{piras2010sphet}
Piras, G. et~al. (2010).
\newblock sphet: Spatial models with heteroskedastic innovations in r.
\newblock {\em Journal of Statistical Software}, 35(1):1--21.

\bibitem[Qu and Lee, 2015]{qu2015estimating}
Qu, X. and Lee, L.-f. (2015).
\newblock Estimating a spatial autoregressive model with an endogenous spatial
  weight matrix.
\newblock {\em Journal of Econometrics}, 184(2):209--232.

\bibitem[Qu et~al., 2016]{qu2016instrumental}
Qu, X., Wang, X., and Lee, L.-f. (2016).
\newblock Instrumental variable estimation of a spatial dynamic panel model
  with endogenous spatial weights when t is small.
\newblock {\em The Econometrics Journal}, 19(3):261--290.

\bibitem[Reich et~al., 2020]{reich2020review}
Reich, B.~J., Yang, S., Guan, Y., Giffin, A.~B., Miller, M.~J., and Rappold,
  A.~G. (2020).
\newblock A review of spatial causal inference methods for environmental and
  epidemiological applications.
\newblock {\em arXiv preprint arXiv:2007.02714}.

\bibitem[Ribeiro~Jr. and Diggle, 2001]{geoR}
Ribeiro~Jr., P. and Diggle, P. (2001).
\newblock {geoR}: a package for geostatistical analysis.
\newblock {\em R-NEWS}, 1(2):15--18.

\bibitem[Rubin, 1974]{rubin1974estimating}
Rubin, D.~B. (1974).
\newblock {Estimating causal effects of treatments in randomized and
  nonrandomized studies}.
\newblock {\em {Journal of Educational Psychology}}, 66(5):688.

\bibitem[Schnell and Papadogeorgou, 2019]{schnell2019mitigating}
Schnell, P. and Papadogeorgou, G. (2019).
\newblock Mitigating unobserved spatial confounding bias with mixed models.
\newblock {\em arXiv preprint arXiv:1907.12150}.

\bibitem[Su and Yang, 2007]{su2007instrumental}
Su, L. and Yang, Z. (2007).
\newblock Instrumental variable quantile estimation of spatial autoregressive
  models.
\newblock {\em Development Economics Working Papers, East Asian Bureau of
  Economic Research}, (22476).

\bibitem[Thaden and Kneib, 2018]{thaden2018structural}
Thaden, H. and Kneib, T. (2018).
\newblock Structural equation models for dealing with spatial confounding.
\newblock {\em The American Statistician}, 72(3):239--252.

\bibitem[Theil, 1958]{theil1958economic}
Theil, H. (1958).
\newblock {\em Economic Forecasts and Policy, Section 6.2. 4}.
\newblock North Holland Publishing Co., Amsterdam.

\bibitem[{US EPA Office of Research and Development},
  2020]{us_epa_office_of_research_and_developmen_2020_4081737}
{US EPA Office of Research and Development} (2020).
\newblock {CMAQ}.
\newblock {https://github.com/USEPA/CMAQ}.

\bibitem[Ver~Hoef et~al., 2018]{ver2018relationship}
Ver~Hoef, J.~M., Hanks, E.~M., and Hooten, M.~B. (2018).
\newblock On the relationship between conditional (car) and simultaneous (sar)
  autoregressive models.
\newblock {\em Spatial statistics}, 25:68--85.

\bibitem[Wright, 1928]{wright1928tariff}
Wright, P.~G. (1928).
\newblock {\em Tariff on animal and vegetable oils}.
\newblock Macmillan Company, New York.

\bibitem[Wyatt et~al., 2020]{wyatt2020annual}
Wyatt, L.~H., Peterson, G. C.~L., Wade, T.~J., Neas, L.~M., and Rappold, A.~G.
  (2020).
\newblock {Annual PM2. 5 and cardiovascular mortality rate data: Trends
  modified by county socioeconomic status in 2,132 US counties}.
\newblock {\em Data in brief}.

\end{thebibliography}

\section*{Acknowledgments}
This work was partially supported by the National Institutes of Health (R01ES031651-01,R01ES027892-01).

\section*{Disclaimer: The views expressed in this manuscript are those of the individual authors and do not necessarily reflect the views and policies of the U.S. Environmental Protection Agency. Mention of trade names or commercial products does not constitute endorsement or recommendation for use.}
\newpage
\section*{Appendix:  Extended sensitivity analysis for local and spatial IVs}\label{extendedSensitivityAppendix}

\begin{figure}[h]
    \centering
    \includegraphics[width=.9\textwidth]{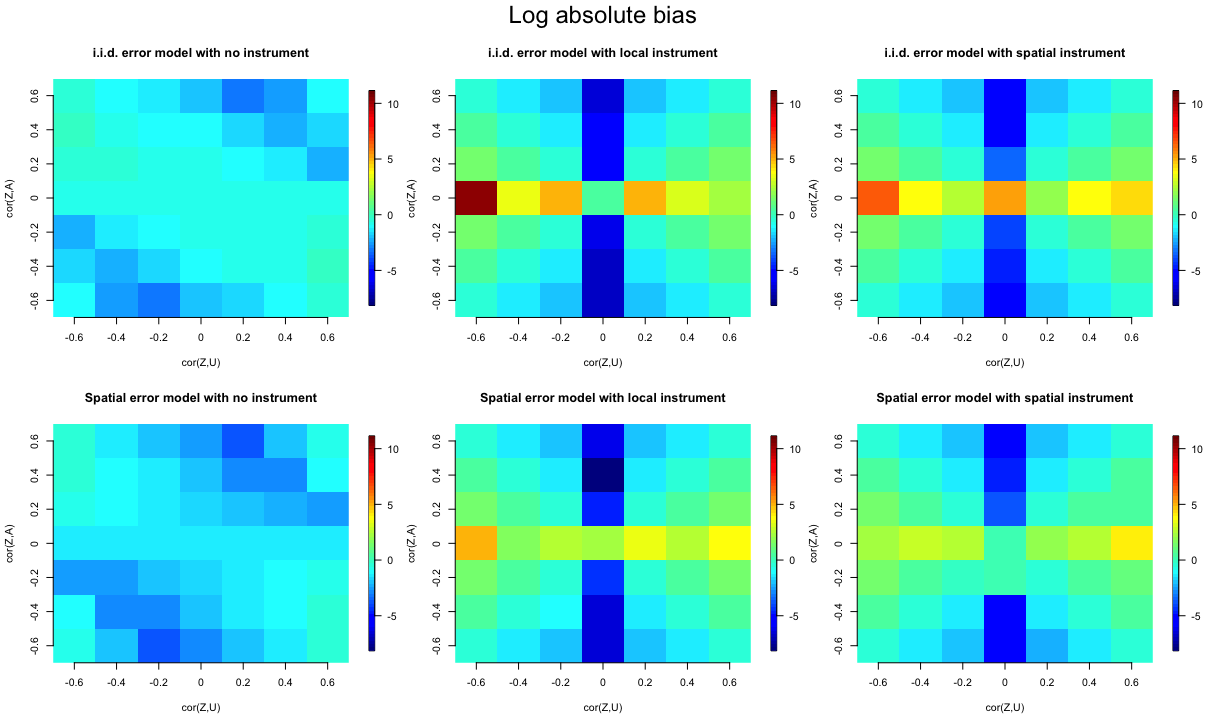}
    \caption{\textbf{Log absolute bias across cor$(Z,U)$ and cor$(Z,A)$ for each model and instrument.} The top row corresponds to the i.i.d.~error models; the bottom row the spatial error models. Within each plot the horizontal axis corresponds to cor$(Z,U)$; the vertical axis corresponds to cor$(Z,A)$.}
    \label{fig:logAbsBias}
\end{figure}

\begin{figure}[h]
    \centering
    \includegraphics[width=.8\textwidth]{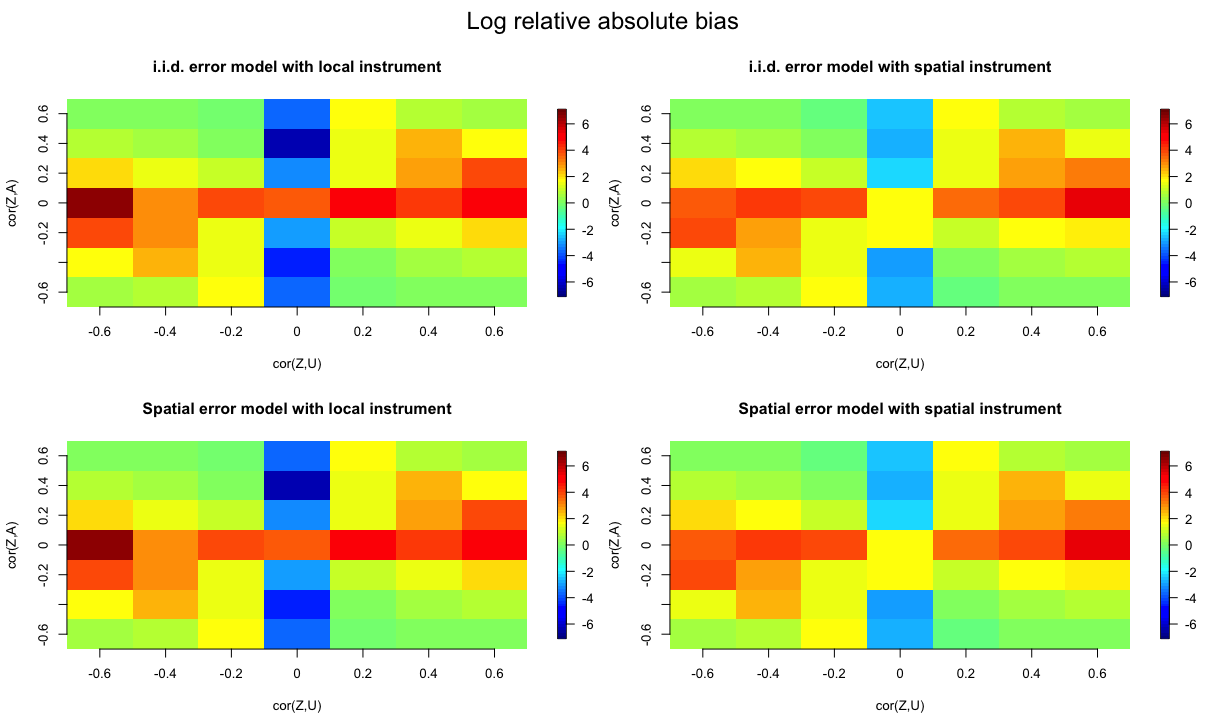}
    \caption{\textbf{Log relative absolute bias across cor$(Z,U)$ and cor$(Z,A)$ for each model and instrument.} The top row corresponds to the i.i.d.~error models; the bottom row the spatial error models. Within each plot the horizontal axis corresponds to cor$(Z,U)$; the vertical axis corresponds to cor$(Z,A)$. The i.i.d.~error models are compared to the no instrument i.i.d.~model; the spatial error models are compared to the no instrument spatial error model.  Log relative improvement in bias for a given model is calculated $\log ( |bias_{\text{model}}| / |bias_{\text{no~IV~model}}| )$.  Thus, values less than one show improvement from the no instrument model; values greater than one show worse performance.}
    \label{fig:logRelAbsBias}
\end{figure}

\begin{figure}[h]
    \centering
    \includegraphics[width=.9\textwidth]{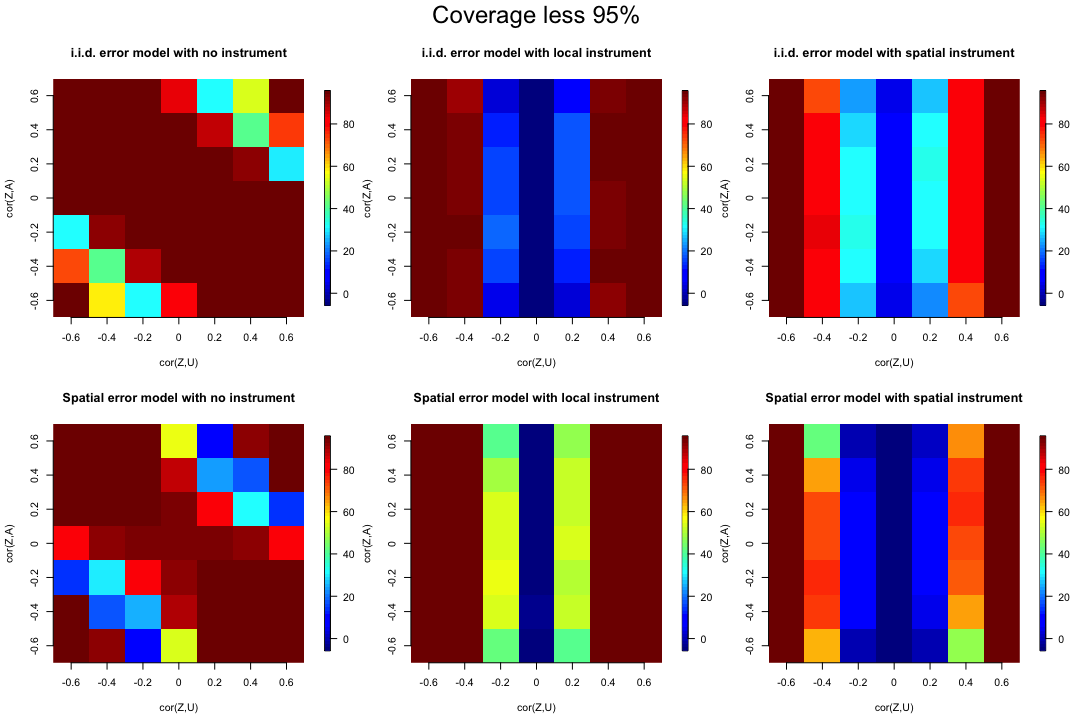}
    \caption{\textbf{Coverage for each model.} The display shows $(95~- $ coverage \%). Therefore, values of 0 correspond to exact 95\% coverage; values below zero show conservative coverage (95\%-100\%); values above zero show anti-conservative coverage (0\%-95\%). The top row corresponds to the i.i.d.~error models; the bottom row the spatial error models. Within each plot the horizontal axis corresponds to cor$(Z,U)$; the vertical axis corresponds to cor$(Z,A)$. }
    \label{fig:coverageLess95}
\end{figure}

This appendix provides an extension of the sensitivity analysis in Section \ref{ss3:squareSim}. The performance of spatial versus local instrumental variables are explored as both the IV strength (cor$(Z,A)$) and the IV validity (cor$(Z,U)$) are varied between $-0.6$ and $0.6$. 
Figure \ref{fig:logAbsBias} examines the log absolute bias: $\log\{ |\hat{\delta} - \delta| \}$ over these ranges. 
Figure \ref{fig:logRelAbsBias} shows these log absolute bias results relative to the baseline no-instrument models. 
Finally, Figure \ref{fig:coverageLess95} explores the coverage given by these models.

\end{document}